\documentclass[letterpaper]{article} %
\usepackage{aaai2026} %
\usepackage{times} %
\usepackage{helvet} %
\usepackage{courier} %
\usepackage[hyphens]{url} %
\usepackage{graphicx} %
\usepackage{natbib} %
\usepackage{caption} %
\usepackage{booktabs}
\usepackage{multirow}
\usepackage{tabularx}
\usepackage{array}
\usepackage{hyperref}
\title{The Promises and Limitations of Evaluation-Based AI Governance:\\ Evidence from Expert Interviews}
\author{
    Lily Stelling\textsuperscript{\rm 1},
    Jacob Charnock\textsuperscript{\rm 1},
    Kevin L. Wei\textsuperscript{\rm 2},
    David Krueger\textsuperscript{\rm 3}
}
\affiliations{
    \textsuperscript{\rm 1}MATS Research\\
    \textsuperscript{\rm 2}Oxford Martin AI Governance Initiative\\
    \textsuperscript{\rm 3}Mila, Universit\'{e} de Montr\'{e}al\\
    hi@stelling.ai
}

\begin{document}
\setcounter{secnumdepth}{2}
\maketitle

\begin{abstract}
We conducted 37 semi-structured expert interviews to examine how frontier AI evaluations reduce societal-level AI risks. Evaluations have become central to AI risk governance, despite their shortcomings being widely critiqued. Still, the case for how they reduce risks in practice remains largely implicit. Drawing on expert accounts, we reconstruct this case by identifying the mechanisms through which evaluations are expected to reduce risk, the assumptions underpinning those mechanisms, and the challenges those assumptions face. Our findings suggest that `evaluation-based governance' could contain foundational gaps which undermine risk-reduction strategies that depend on evaluations. Using thematic analysis, we identified three key pathways through which evaluations could reduce risk: by (i) making societal-level risks concrete to decision-makers and the public (raised in 18 out of 37 interviews), (ii) serving as a basis for governance commitments and regulation (18), and (iii) incentivizing a better risk culture within AI companies (16). However, these pathways often depended on shared assumptions that participants themselves doubted. The main challenges included, in order of frequency: (i) diminishing informativeness of evaluation results as model capabilities improve (21), (ii) measurement limitations that may be inherent to the evaluation paradigm (18), (iii) a lack of guarantees that evaluation results will be acted upon by companies or governments (16), and (iv) the degree of control and discretion AI companies retain over evaluation design, disclosure, and response (12). Participants also noted that evaluation-based governance could prove counterproductive, by functioning as a permission structure for developers' continued capability improvement (4). We finish by discussing the implications of our findings for the evaluation-based governance paradigm, and the conditions that would improve its prospects.
\end{abstract}

\section{Introduction and Related Work}
 \label{sec:related-work}
The development and deployment of frontier AI systems can pose societal-level risks, including risks to democracy through mis- and disinformation; to national security through biosecurity, cybersecurity, and misalignment threats; and to economic organization through labor displacement and global inequality \citep{Bengio2026, Goldstein2023, Burton2025, DrexelWithers2024, Eloundou2024}. As a response, frontier AI developers have embraced `evaluation-based governance': utilizing measurements of AI behaviors to identify AI capabilities, performance, vulnerabilities, or impacts \citep{Bengio2026, Shevlane2023}, as a basis for assessing and mitigating risks from frontier AI.
 
Evaluations have long been used to assess AI systems for bias, fairness, and other social harms \citep{Buolamwini2018, Selbst2019}. However, evaluations of dangerous capabilities now play a distinct role. Emerging regulatory frameworks such as the EU AI Act \citep{EUAIA2024}, California's SB-53 \citep{california2025sb53}, and the Frontier AI Safety Commitments \citep{DSIT2024} rely on dangerous capability evaluations as the empirical basis for risk classification and mitigation triggers.

The success of these governance regimes depends on reliable, accurate, and trustworthy AI evaluations. However, evaluations suffer from well-documented limitations, and a growing body of work questions whether AI evaluations can 
bear the governance weight placed on them. Foundational 
critiques have shown that general-purpose benchmarks suffer 
from construct validity failures, measuring proxy tasks that 
may not correspond to the real-world phenomena they claim to 
capture \citep{Raji2021,JacobsWallach2021, 
Eriksson2025}. Benchmark culture can also encode values 
that systematically centralize power around those who design 
the tests \citep{BirhaneKalluriCard2022}. These concerns, 
originally developed in the context of fairness and bias 
evaluations \citep{Buolamwini2018, Selbst2019, Raji2021, 
Ibrahim2024}, may extend to dangerous capability evaluations, 
which suffer from additional challenges: rapid benchmark 
saturation \citep{Reuel2024BetterBench, Kiela2021, AnthropicEvals2026}, 
sensitivity to elicitation choices \citep{Casper2024, 
Burden2024}, and omission of contextual information from 
developers' published reports \citep{Staufer2025, 
Bommasani2024}. Evaluations for dangerous capabilities can 
also be constructed in forms that may serve developers' 
interests \citep{Eriksson2025, Ren2024, HardyReuel2024}.
 
Many evaluations are built, run, and publicized by frontier 
AI developers themselves, potentially mirroring the 
agenda-setting form of regulatory capture documented in AI 
policy more broadly \citep{Wei2024, AbdallaAbdalla2021, 
Birhane2026}. Governance frameworks draw heavily on 
approaches that developers originated \citep{EUAIA2024, 
DSIT2024, Anderljung2023, Buhl2025}, which some suggest were 
adopted partly because labs were already using evaluations 
for capability improvement and marketing 
\citep{Ren2024, BirhaneKalluriCard2022}. More broadly, corporate logics 
can drive governance work toward ``checklists, procedures, 
and evaluative metrics'' that convert political questions 
into tractable engineering problems 
\citep{MetcalfMossBoyd2019}, and AI principles alone may be 
insufficient without the institutional infrastructure to 
enforce them \citep{Mittelstadt2019}. Recent work has 
taxonomized how firms can walk the line between legal 
compliance and substantive avoidance under risk-based AI 
regulation \citep{Yew2025}, raising questions about whether 
evaluation-based governance could be similarly circumvented.

Interview-based studies of responsible AI practitioners have 
documented how procedural governance instruments function in 
practice. Evaluations and checklists only operate as 
governance when integrated with organizational workflows and 
incentives \citep{Madaio2020}; practitioners often face 
structural barriers including product-launch supremacy and 
difficulty quantifying ethical concerns 
\citep{Ali2023, Rakova2021}; and model-level evaluations can 
fail as governance when the actors being evaluated lack the 
power or incentives to act on findings 
\citep{WidderNafus2023}. Field scans of the algorithmic audit 
ecosystem have identified access, independence, and 
enforceability as persistent bottlenecks 
\citep{CostanzaChockRajiBuolamwini2022, Friedler2023}. 
However, most of this work examines fairness and bias 
evaluations within industry settings. How evaluation-based 
governance functions for frontier AI risks, where the stakes 
extend to national security and societal-level harms, and 
where the evaluation paradigm is now embedded in regulation, 
remains largely unexamined. The closest methodological 
precedent is \citet{Wei2024}, a 17-interview study of US AI 
policy experts on regulatory capture dynamics. However, it remains understudied how and under what conditions evaluation-based governance translates into reduced risk.
 
This article addresses that gap. Through 37 semi-structured interviews with AI governance experts, technical researchers, and evaluation developers, we surface the tensions and fragilities of evaluation-based governance, pinpoint where core assumptions may break down, and highlight what targeted interventions could strengthen them. Using thematic analysis, we identify three key pathways through which evaluations could reduce risk: by (1) making societal-level risks concrete to decision-makers and the public, (2) serving as a basis for governance commitments and regulation, and (3) incentivizing a better safety culture within AI companies. However, participants widely questioned the assumptions 
these pathways depend on, and several argued that evaluation-based governance could prove counterproductive, by providing a false appearance of responsible action.
 
Our findings do not suggest that evaluations cannot support AI governance, or that evaluation-based governance frameworks will necessarily fail at reducing risks. Rather, they help identify the institutional conditions that could be needed for evaluations to reliably reduce risks from AI. We conclude by discussing recent developments that may have reinforced many participants' concerns, and the conditions under which evaluation-based governance could be strengthened.

\section{Methodology}
 
We conducted 37 semi-structured interviews with experts on AI evaluations and governance, and the mechanisms for how evaluation-based governance could reduce risk (`theories of change'). A semi-structured method is appropriate because the strategic reasoning and institutional knowledge behind evaluation-based governance is rarely articulated in published work, meaning theories of change and their assumptions may be best surfaced through open-ended expert discussion \citep{Soest2023}. Interviews were anonymized to encourage candid reflection on governance strategies and institutional dynamics.
 
\subsection{Participant Selection}
 
We used purposive and snowball sampling to identify participants based on their professional expertise in AI evaluations and governance. We contacted approximately 140 experts and conducted 37 interviews across four stakeholder groups: AI governance researchers, evaluation developers, technical AI researchers, and AI advisors in government. After 27 interviews, no new themes emerged, indicating saturation \citep{Saunders2018}. Participants were based primarily in the United States and the United Kingdom, with some in the European Union. The sample includes participants currently or formerly affiliated with frontier AI companies, government agencies, nonprofits, and academia. Participant backgrounds are described in Table 2 in the appendix.\footnote{Given the small size of our population of interest, detailed demographics are not reported to protect participant confidentiality.}
 
\subsection{Interview Protocol}
 
All interviews were conducted online by the first author between February and April 2025, lasting 60--70 minutes each. The first author described the study goals, outlined confidentiality practices, and obtained verbal informed consent. Participants received a participant information sheet in advance, were informed that participation was voluntary and they could decline any question or withdraw at any time, and were told the anonymization process in full.
 
The protocol was structured around four sections. The first covered participants' background and definitions of AI evaluations. The second explored theories of change for how evaluations could reduce societal-level AI risks, what non-technical conditions participants considered necessary for success, and how plausible they judged those conditions to be.\footnote{We focus on dangerous capability and societal-level risk evaluations because of their role in regulation, though many challenges we document may translate to other evaluation contexts.} The third addressed limitations and criticisms, including scenarios where accurate evaluations could fail to prevent harm, as well as underappreciated negative effects. Technical researchers were asked additional questions on the state of evaluation science, and the difficulty of control and propensity evaluations. The protocol was refined after three pilot interviews. The full protocol is in Appendix C.
 
\begin{figure}[t]
\centering
\includegraphics[width=\columnwidth]{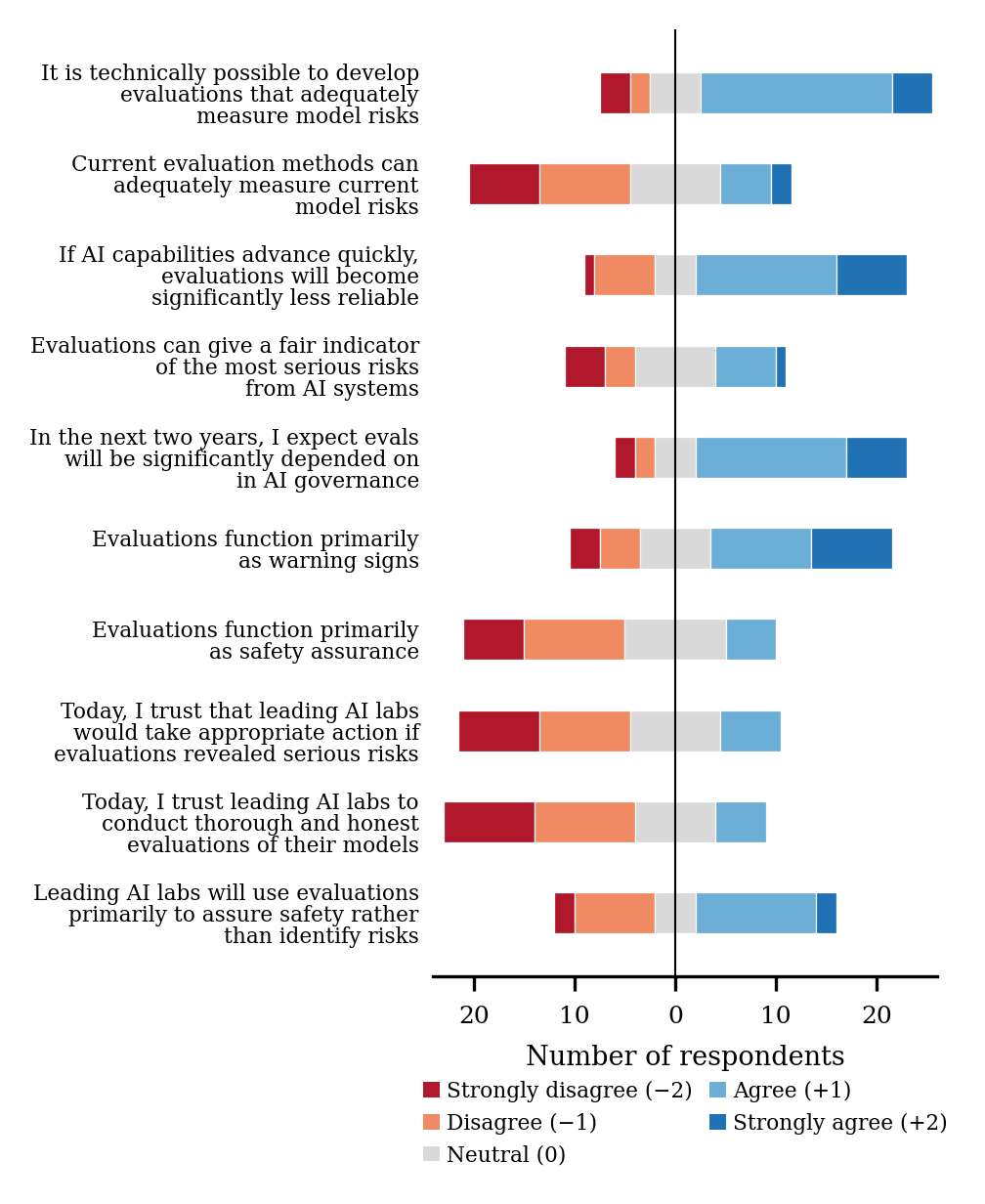}
\caption{Participant responses to Likert-scale questions ($-2$ = strongly disagree, $+2$ = strongly agree). Bars show number of respondents in each category, centered on neutral. Not all participants answered every question.}
\label{fig:likert}
\end{figure}

The interviews also included a short quantitative component. At the end of each interview, participants responded to Likert-scale questions (five-point scale, $-2$ to $+2$) covering trust in developers to conduct evaluations honestly, trust in developers to act on concerning results, and expectations about evaluation reliability as capabilities advance. 33 of 37 participants completed these questions. Results are reported in Figure~\ref{fig:likert}.

\subsection{Data Analysis} 
Interviews were transcribed using a local instance of OpenAI's Whisper model, then anonymized following a speaker-identifying-only principle \citep{SaundersKitzingerKitzinger2015}. At no point were recordings or identifiable data input to AI systems other than the local transcription tool.
Transcripts were analyzed using hybrid thematic analysis \citep{FeredayMuirCochrane2006, King2012}, combining 30 deductive codes derived from the interview protocol and existing literature with inductive codes that emerged during coding. The final codebook contains 250 codes, of which 220 were emergent. Codes were organized into families including theories of change, limitations, critiques, assumptions, strategies, and valence codes of optimism or pessimism. To calibrate the codebook, both coders independently coded three transcripts and compared results. Disagreements were discussed and code definitions revised. A fourth calibration transcript was then independently coded by both coders, yielding high agreement. The remaining 33 transcripts were divided between the two coders, with joint discussion on approximately five transcripts where either coder was uncertain. We followed a negotiated agreement approach \citep{Campbell2013}.\footnote{Because our coding approach was quote-based rather than segment-based, unit-by-unit inter-rater reliability metrics such as Cohen's kappa were not appropriate \citep{Campbell2013}.}
 
To identify the assumptions underlying each theory of change (RQ2), we drew on three sources: assumptions participants explicitly articulated, latent assumptions identified through interpretive coding \citep{Saldana2021, BraunClarke2006}, and assumptions surfaced by interpretation of participants' criticisms. The first author produced an initial mapping; the second coder independently reviewed it, and disagreements were resolved through discussion.
\section{RQ1: What Mechanisms Do AI Evaluation Experts Identify for How Evaluation-Based AI Governance Could Reduce Societal-Level Risk?}
 
\subsection{Evaluations Make Risks Concrete}
 
The most widely raised theory of change cast evaluations as a ``wake-up'' signal, making AI risks concrete to decision-makers who might otherwise remain unaware of what frontier models are capable of (18 of 37 participants). The reasoning was as follows.
 
\paragraph{Advance warning.}
Evaluations could detect dangerous capabilities before models are publicly released, giving developers and policymakers time to implement safeguards. The idea is that results could reveal something alarming enough to update decision-makers' understanding of risk, generating political will to act before harm occurs. Internal evaluation results can let developers and those with model access ``see capabilities up to a year before these capabilities would be made public,'' meaning concerning findings could reach national security officials before public release (US24). This advance warning may be necessary because ``waiting for large real-world impacts is probably too late,'' as developers may ``build [artificial general intelligence] in a lab [...]  and use it to automate internal research,'' meaning the effects of advanced capabilities could arrive without any public release to serve as a warning (US27). UK18 described this mechanism more generally as ``mov[ing] information earlier in time'' so that institutions have more time to prepare.
 
\paragraph{Building political momentum.}
Evaluations could generate momentum for stronger regulatory actions by building consensus for the requirement of if-then commitments (i.e.\ pre-agreed commitments specifying that if a model crosses a given capability threshold, then specific safeguards must be implemented to appropriately reduce risks), red lines (i.e.\ pre-agreed capability thresholds that should not be crossed), and/or regulation: ``If you can start getting some traction with early signs or early results from evaluations, you can get people to buy more into the kind of if-then framing'' (US26). Evaluations could also increase the perceived urgency of AI risk among governments, leading them to ``invest time and energy and resources into solving it'' (P15).
 
\paragraph{Legible evidence.}
Evaluations may be well-suited for providing a wake-up 
signal because they can produce legible empirical evidence 
that conveys risks more effectively than alternatives. I14 
framed this as a departure from earlier AI safety work: ``a 
lot of concerns around AI safety have been pretty abstract 
and speculative [\ldots{}] whereas evaluations are much more 
concrete and rigorous, you can measure things, get numbers, 
plot trend lines on graphs.'' Several participants drew 
explicit contrasts with demonstrations (i.e.\ curated 
examples of model behavior shown to an audience), arguing 
that evaluation results may be harder to dismiss because 
audiences perceive them as more rigorous. ``Lots of people 
have this pretty reasonable reaction to demos where they're 
like, oh, I guess the thing has been cherry picked to be 
scary,'' whereas ``people's intuitive reaction to evaluations 
is much more like, oh, the fancy Stanford professor did 
science on the model'' (P15).

\subsection{Evaluations Form the Basis for Governance Commitments and Regulation}
 
A second widely held mechanism suggested that evaluations could provide the empirical foundation for governance frameworks that reduce AI risks, including voluntary lab commitments and formal regulation (18 of 37 participants).
 
\paragraph{Ground commitments.}
Evaluations could provide empirical evidence to operationalize concrete commitments. Developers can set measurable thresholds such that ``if the model exceeds `X' score on a single benchmark, then we will make sure that we have such and such level of safeguards ready in time'' (US37). An even stronger version involves tying evaluation results to pausing development entirely \citep{AlagaSchuett2023}, where developers can run evaluations ``early and frequently'' to produce ``early warning signs of potential risks'' that allow them to ``forecast how we might expect capabilities and risk to evolve'' and, if necessary, ``pause development if it's a really extreme case'' (US23).
 
\paragraph{Establish red lines.}
By identifying current capabilities and alerting when more dangerous ones might emerge, evaluations could let stakeholders agree on which capability thresholds should not be crossed. ``How close we are to those catastrophic levels'' becomes trackable (US28), making it possible to ``agree in advance'' on which capabilities are unacceptable ``and at what level they would occur'' (US29). Measurability may be a requisite: ``for the red line story to go well, you probably need a good eval that measures whatever the red line is'' (A6).
 
\paragraph{Raising the industry floor.}
Companies adopting evaluation-based governance could create reputational pressures for others to follow. Evaluations may play ``a useful role in both hard and soft law, with companies feeling peer pressure to run dangerous capability evals'' (A1). ``Other companies feel a little bit of peer pressure from, let's say, Anthropic, OpenAI, Google DeepMind to do evaluations for dangerous capabilities and try to set up mitigations in place'' (US37). Public commitments could even make formal regulation more feasible by maturing industry best practice, with evaluation-based governance serving ``first as industry standards'' and then ``getting into legislation'' (US34).

\subsection{Evaluations Shape Safety Incentives Within AI Companies}
 
A third mechanism is that evaluations could positively influence the incentives facing AI developers, pushing them toward better safety practices (16 of 37 participants).
 
\paragraph{Refine safeguards.}
Frontier developers can use evaluations to identify new vulnerabilities and test whether existing safety measures (`safeguards') work. Evaluations can be ``used to improve mitigation methods [\ldots{}]  as you iterate against them'' (UK20), and ``you can see how well your alignment techniques are working using the evals'' (US27). In some cases, this may have immediate effect as developers ``see their evaluation [scores] have gone up and then [implement] some mitigations, like doing RL to prevent [models] from giving biological weapons help'' (US27).
 
\paragraph{Improve understanding of risk domains.}
Building evaluations requires developers to allocate staff and resources to specific risks they might otherwise neglect. ``If Anthropic were not doing any CBRN evals,\footnote{Chemical, biological, radiological, and nuclear capability evaluations.} same for OpenAI, then they would not be thinking as much about getting some safeguards ready in  place'' (US37). US35 pointed to multi-agent collusion as a domain where this has not yet happened: ``we have no evaluations for collusion and companies don't even run them as far as we know,'' even though ``there might be risks of AI systems basically conspiring with each other.'' Constructing evaluations can also surface unexpected failure modes by forcing researchers to think through threat models they had not previously considered.
 
\paragraph{Establish accountability through documentation.}
Evaluations could provide transparency about model capabilities, establishing accountability. Public commitments may create reputational stakes, where ``if a company publishes something like [a public commitment] and then they don't stick to it [\ldots{}]  [damage] to their reputation [keeps] them in line'' (A2). Evaluations could also give more risk-minded employees ``another lever to pull, to try to develop really clear arguments, to try to kind of retain some power to say, `slow down' when we need to slow down'' (A3). Accountability could also come from enterprise customers, where ``if you just pull a CEO saying, `hey, are you considering working with [frontier AI developer]? Here's some scheming that came from their latest results','' customers may inquire further of developers (EU10). Documentation could create a record developers ``can be held accountable to,'' whereas absent evidence they can ``always claim ignorance if there is an accident'' (US25). These records may accumulate such that ``the evidence [is] just piling up and becoming so egregious,'' to potentially be utilized in legal contexts (US27). Documenting evaluation results could also ``add time, and [\ldots{}]  add additional eyes and scrutiny'' that matures safety culture, as developers may have ``some fear that problems will be discovered'' (A1). 
 
\section{RQ2: What Assumptions Underpin These Theories of Change?}

These three pathways appear to share a common causal logic: evaluations produce reliable information, that information reaches the right people, and those people respond in ways that reduce risk.  Table~\ref{tab:assumptions} presents 
the full set of assumptions we identified, organized into 
foundational assumptions that would need to hold for each identified 
pathway to succeed, and pathway-specific assumptions unique to 
each theory of change (see Appendix B for the 
latter). 
 
 The sections that follow examine where participants identified these assumptions as most likely to fail.

\begin{table*}[t]
\centering
\small
\begin{tabularx}{0.9\textwidth}{>{\raggedright\arraybackslash}p{2.8cm} >{\raggedright\arraybackslash}p{4.5cm} X}
\toprule
\textbf{Family} & \textbf{Assumption} & \textbf{Key challenges identified by participants} \\
\midrule
\multirow{4}{*}{\parbox{2.8cm}{Evaluations can produce a trustworthy signal}}
& Evaluators can elicit capabilities at or near the ceiling of model performance
& Elicitation may be highly sensitive to prompting, scaffolding, and tool access (\S\ref{sec:measurement}); developers could have structural advantages over external evaluators in elicitation quality (\S\ref{sec:labcontrol}) \\
\cmidrule(l){2-3}
& Evaluation performance corresponds to real-world risk (i.e.\ construct validity) 
& Proxy tasks may not map onto real-world threat scenarios (\S\ref{sec:measurement}); stakeholders may not be able to clearly interpret what evaluation results signify (\S\ref{sec:labcontrol}) \\
\cmidrule(l){2-3}
& Evaluations chosen cover the relevant threat surface, including emergent risk domains
& Evaluations may neglect more complex risk surfaces, such as multi-agent interactions, longer time horizons, or simply unknown unknowns (\S\ref{sec:diminishing}) \\
\cmidrule(l){2-3}
& Models do not strategically underperform during evaluation
& Models could develop sufficient `evaluation awareness' to detect evaluation conditions and alter behavior, undermining evaluations' informativeness (\S\ref{sec:diminishing}) \\
\midrule
\parbox{2.8cm}{Evaluations are run fairly and honestly}
& Developers conduct evaluations with sufficient effort, rigor, and frequency
& Developers are not incentivized to conduct evaluations with sufficient rigor, and control what is tested, how results are reported, and which findings are disclosed (\S\ref{sec:labcontrol}) \\
\midrule
\parbox{2.8cm}{Evaluations produce trustworthy signal over time}
& Evaluation methodology keeps pace with the rate of model development
& Benchmarks may saturate faster than replacements can be developed (\S\ref{sec:diminishing}); evaluating increasingly complex agentic behaviors could outstrip available methodology (\S\ref{sec:diminishing}); evaluation awareness could worsen over time (\S\ref{sec:diminishing}) \\
\midrule
\parbox{2.8cm}{Results reach decision-makers}
& Evaluation results are communicated to relevant audiences in an interpretable form
& Companies can engage in selective disclosure and reporting (\S\ref{sec:labcontrol}) \\
\midrule
\multirow{2}{*}{\parbox{2.8cm}{Results are acted upon}}
& Appropriate responses to mitigate risk are available when needed
& Mitigations could lag behind capability development (\S\ref{sec:noaction}); developers may resist building safeguards before problems emerge (\S\ref{sec:noaction}); the ecosystem could be overfocused on detection relative to response (\S\ref{sec:noaction}) \\
\cmidrule(l){2-3}
& Sufficient political will exists to create and enforce regulatory responses to evaluation results
& Voluntary commitments may lack enforcement (\S\ref{sec:noaction}); commercial pressure could incentivize favorable interpretation of results (\S\ref{sec:noaction}); political will for regulation could be absent or insufficient (\S\ref{sec:noaction}); commercial and racing pressures could override safety considerations (\S\ref{sec:noaction}) \\
\bottomrule
\end{tabularx}
\caption{Foundational assumptions underlying key pathways for evaluation-based governance to reduce risk.}
\label{tab:assumptions}
\end{table*}

\section{RQ3: What Criticisms Do Experts Raise That Challenge These Assumptions?}

\subsection{Measurement Limitations}
\label{sec:measurement}

AI evaluations face well-documented measurement limitations 
(see Section~1). As one participant summarized, ``just 
because your model passes all your evals, it doesn't mean 
it's necessarily safe. We don't have any bounds or 
guarantees on that'' (A5). Another concurred: ``no one's 
satisfied with the evaluations that we currently have. 
Everyone's like, oh man, is this even measuring the right 
thing?'' (I12). Below, we focus on structural and 
institutional challenges that participants identified as 
potentially undermining the governance role evaluations are 
asked to play.

\subsection{No Reliable Path From Information to Action}
\label{sec:noaction}
 
A key criticism was that even if evaluations surface valuable information, no clear institutional processes may exist to translate results into appropriate responses: ``if an [evaluation result] comes out and looks really scary, it's not clear that anything happens'' (P15). As a result, ``[evaluation-based governance] can totally just fail because people don't listen to [evaluations]'' (A1).
 
\paragraph{Voluntary commitments may lack enforcement mechanisms.}
Participants widely doubted that voluntary commitments would hold under pressure (A2, A6, EU10, P15, US24, US28, US35, US37). As A2 noted, ``you don't have to respect [your commitments] at all. You can kind of just change it at any time, as multiple companies have done. And no one really cares.'' Companies could invoke `exit clauses' (i.e.\ provisions in 
voluntary commitments that allow developers to deviate from 
stated safeguards under specified conditions) when results become inconvenient: ``the thing I imagine happening [\ldots{}]  is that the lab is like, oh yeah, it triggers XYZ in our [voluntary commitments], but actually we're going to trigger our exit clause [\ldots] [and] that seems reasonably likely'' (P15).
 
\paragraph{Companies may be incentivized to bypass safety considerations.}
Commercial pressure could create incentives to interpret evaluation results favorably. This may be especially acute in the AI industry, as one participant highlighted: ``there was the Volkswagen scandal [\ldots{}]  what is the economic incentive of that compared to the hundreds of billions we're now talking about? It's insane how high the incentives are to be below a well-defined threshold'' (EU8). Similarly, motivated reasoning could shape individual deployment decisions, where ``it wouldn't surprise me if people end up finding, even subconsciously, ways to make it across that threshold [\ldots{}]  they would just be like, oh well, we have this mitigation now so that brings it down to this level'' (I14). The dual-use nature of many capabilities also provides readily available justifications, as ``the same capability can also 
be used defensively\ldots{} and so I think there's a certain 
way in which people kind of justify the dangerous capabilities 
by flipping it to the defensive version of them'' (I14). Put starkly, without external oversight, ``why would I trust a 
profit-driven organisation to follow something that is 
against their profit-driving mechanism?'' (EU10).
 
\paragraph{Political will to create enforcement mechanisms may be absent or insufficient.}
Participants noted that the regulation needed to give evaluation-based governance teeth may be unlikely to arrive in time. As UK20 noted, ``the chance that we get governance that is sufficient in my mind is already gone [\ldots{}]  It's only a question of risk reduction at this point.'' The US political environment was raised extensively: ``under the current administration, it seems unlikely that they would require evaluations [of developers]'' (US22). Similarly, ``in order to really mitigate the risks, probably you're going to need some kind of third-party auditing [\ldots{}]  and I think now with the changes in the political environment that's probably looking less likely'' (I14). Even where political will does exist, participants were unsure if governments ``actually have the state capacity, decision-making ability, ability to align on a course of action'' amidst rapid capability progress (UK17).
 
\paragraph{Racing dynamics may override safety considerations.}
Competitive dynamics between companies and nations could 
create structural pressure to continue development, 
regardless of evaluation results (A6, US24, US37, UK20, A2). 
As US24 noted, ``People are just not going to abide by their 
commitments. They're going to be like sorry this other lab is 
racing too much [\ldots] or China, look at the Chinese model, 
so scary.'' Racing dynamics could also operate at the 
government level, with nations deprioritizing responsible 
deployment under competitive geopolitical pressure (A6).
\paragraph{Mitigations may not be ready even when evaluations signal their requirement.}
Even where evaluations function well and commitments are honored, weak mitigations could undermine risk reduction, as ``probably the most fragile assumption [is] whether the safeguards are actually ready in time'' (US37). One participant described this as a possible ecosystem failure of overfocusing on detection at the expense of response (US26). 
 The difficulty scales with the severity of the risk. US37 
framed this as the ``crucial question'': ``if it's fairly 
easy, then existing frontier safety policies could be 
sufficient. If it's very hard, if it requires months of 
pausing and lots of coordination between AI labs to avoid an 
[artificial general intelligence] race, I guess we're just 
not on track to do that.''
 
This problem could be compounded by a structural reluctance within labs to invest in preemptive mitigations, which participants attributed to startup culture and a norm of waiting for empirical evidence before building safeguards (I14).
 
These concerns could undermine two foundational assumptions: that institutional mechanisms exist to translate evaluation results into appropriate responses, and that results are acted upon when they reach decision-makers (Table~\ref{tab:assumptions}, ``Results are acted upon'').

\subsection{Frontier Developers Retain Significant Control Over Evaluations}
\label{sec:labcontrol}
 
Even where evaluations produce valuable information, participants identified a translation gap between results and their uptake. ``Most policymakers probably don't have a view [on evaluation-based governance] because it's not even something they've heard much about'' (US34), and ``the underground experience is like, actually, they just don't know [about evaluation results]'' (US30). Even when results reach audiences, they may fail to communicate risk in actionable terms, as ``people agree what the numbers are, but they don't agree what it means'' (US36). This translation gap could be compounded by the degree of control frontier developers retain over the evaluation ecosystem, such as what is measured, how results are reported, and who has access.

\paragraph{Frontier developers control how evaluations are reported.}
Companies' reporting practices vary widely \citep{Bommasani2024, Staufer2025}, and companies hold discretion on what goes into public system cards (A1, A2, US37, US24). Participants noted that ``there's no standardized way to report on things yet'' because ``the science of evaluations is pretty immature'' (A1). This could allow results to be presented in misleading ways, since ``there are so many details that go into how an evaluation is done, how rigorous it was, and what the results all mean'' (A1). 
 
Companies could use the subjectivity of evaluations to ``justify [the model] being safe'' without needing to ``literally lie'' (I14), for instance by arguing a capability threshold was crossed only because the model memorized test items, or that a new mitigation brings the risk level down sufficiently (I14). Developers could also cherry-pick which results to publicize 
without consequence, presenting safety evaluations to ``say 
that it's safe'' when releasing a model, while selecting 
``the benchmark that they can show they have done better than 
others'' when competing on capability (US31). Developers may also deliberately withhold harmful results to protect their reputation, or because disclosure might ``hurt our stock price'' (US24). Running evaluations thoroughly and reporting on them constrains internal capacity, meaning ``it's just not really in their interest to do a good job here'' (A2), leaving ``a lot to be desired in terms of elicitation.''
 
\paragraph{Frontier developers control thresholds and responses.}
Developers retain significant discretion over the thresholds and safeguards that trigger commitments. Frontier safety policies contain ``a lot of subjective wording,'' letting developers ``remain flexible'' about when commitments trigger (US37), and the mapping from eval results to capability thresholds is often ``really ad hoc'' and ``somewhat arbitrary'' (US23). This discretion could be concerning because developers may lack incentives to interpret evaluation results in ways that prioritize safety when these conflict with commercial interests (EU10).
 
Companies may have valid reasons to withhold full evaluation results, including information hazards\footnote{This was a ``live concern on the bio side of things,'' where evaluation design ``by necessity'' compiles dangerous information to test for biological risks (US26).} and benchmark contamination concerns. Nevertheless, several participants were skeptical that developers would inform regulators or policymakers when results warranted concern (e.g.\ US24). Moreover, developers could use the fragility of evaluations to create the narrative that ``evaluations are really hard, and we don't know how to do evaluations well, therefore we can't do regulation'' because ``good evaluations are a precondition for regulation'' (US27).
 
\paragraph{Insufficient access for meaningful external evaluations.}
Interviews suggested that developers may control the access external evaluators receive. Participants noted that if third-party evaluators typically receive only query access, this is ``extremely limiting'' for external testing because ``all the state of the art red teaming tools use white box access'' (A1). As a result, ``the evaluations aren't even really valid [\ldots{}]  for the purposes that people want them for because of this limited access'' (US29). Even where some access is granted, people at developers are ``much better at elicitation than people outside'' because of extensive ``hands-on experience'' with the models (P15). A persistent issue raised was that independent evaluators who depend on developers for access may find it difficult to present critical findings without risking the partnerships their work relies on (A1).

This issue could be compounded by developers continuously improving their models up to release, meaning that even when access is obtained, the version external testers evaluate often differs from what gets released \citep{Charnock2026}. Developers are ``basically always [\ldots{}]  fine-tuning [their] model'' while testers are given access only ``a month or so before'' and are therefore ``not testing the final version inherently'' (UK18).
 
\paragraph{Frontier developers control the evaluations chosen.}
Because frontier companies have discretion over which evaluations they give access to conduct, the evaluation ecosystem could skew toward developers' commercial interests. Publicly available evaluations could serve as a training signal, as ``you can basically just do RL using the evaluation as the signal'' and this ``might actually be the bottleneck'' for further capability improvement (US27). Participants worried about this, as safety performance may coincide with capabilities in general (US29; see Ren et al., 2024). Developers may be ``keen to run'' cheap multiple-choice evaluations because they are ``useful as an internal measure of capability,'' but it could be ``pretty difficult to expect [the companies] to develop the super in-depth methodologies for different types of threat models [\ldots{}]  needed [\ldots{}]  for safety reasons'' (I14).

These dynamics could undermine the assumptions that evaluations are run fairly and honestly, that evaluators can elicit capabilities near the ceiling, and that results reach decision-makers without distortion (Table~\ref{tab:assumptions}).

\subsection{Evaluations May Have Diminishing Returns as Model Capabilities Improve}
\label{sec:diminishing}

\paragraph{Evaluation methodology may not keep pace with model development.}
The speed and complexity of capability progress could outpace evaluation development (US23, US27, EU10, US22, US30, UK18, UK17). As US23 noted, ``there's a super high degree of saturation on all the evaluations, which would indicate that models are advancing much more rapidly than we can advance the evaluations ecosystem'' (US23). As a result, ``the frequency of new evaluations or evaluations updates cannot really match the frequency of model updates'' (EU10). 
 
These problems may be most severe where capabilities advance discontinuously, such as if AI systems become proficient enough at coding and research to accelerate their own development \citep{Davidson2023}. Beyond speed, the \textit{complexity} of behavior that evaluations need to capture could outstrip available methodology (US23, A6, P15, A2, I14). Current evaluations may fail to capture ``interactions between models and systems'' since they evaluate ``one model'' without ``[taking] into consideration its interactions with the real world and interactions with other highly capable models'' (US23). As AI models move toward agentic tasks with tool access and longer time horizons, ``there's not a science of this yet'' to rigorously evaluate for risks in this context (P15). Even evaluations that successfully capture this complexity may be too costly, as they demand ``longer evaluations'' and ``more realistic tasks'' that ``just get really painful to run'' (A2).
 
\paragraph{More capable models may show evaluation awareness.}
Evaluations could become less effective as models develop evaluation awareness (i.e.\ the ability for models to detect when they are being tested rather than deployed; \citealt{Berglund2023, Needham2025}) and alter their behavior accordingly (US27, EU8/9, A6, A2, I14, US24, US28, UK21, UK20, UK19, US35, US36). The clearest case is sandbagging (i.e.\ strategic underperformance on an evaluation), which could diminish an evaluation's informativeness \citep{vanderWeij2024}. Models might ``sandbag or [\ldots{}]  manipulate the results'' (US35), by drawing on ``enough situational awareness to understand [they are] being evaluated'' (US28). One participant framed this as ``an empirical race between evaluation awareness and the evolution of scheming'' (i.e.\ a model pursuing goals misaligned with its developers while appearing compliant): if ``evaluation awareness [of AI models] evolves really strongly [\ldots{}] then you're maybe just screwed'' (A2). This could create a ``false security'' problem where models become ``better at hiding [misaligned behaviors] [\ldots{}]  instead of actually being more aligned'' (A2), making it ``very hard to trust'' evaluation results going forward (I14).

\paragraph{Evaluations may become less effective for safety cases as capabilities improve.}

Several participants argued that evaluations could become less effective for safety cases (i.e.\ reports that make a structured argument, supported by evidence, that a system is safe enough in a given operational context; \citealt{Buhl2025}) as capabilities improve. Evaluations ``become less impactful once we leave the regime where having a low score on a benchmark can tell you that an AI is safe'' (the `inability argument')\footnote{The inability argument is the claim that ``AI systems are too weak to be dangerous'' even under worst-case assumptions \citep{Clymer2024}.} (US35). ``A lot of [evals] become much less useful'' once ``models are [\ldots] above human expert level'' because ``the assumption [becomes] that they are capable and that safety relies on other mechanisms'' (UK19). One participant argued that evaluations could become redundant once models are clearly dangerous, because at that point ``even the best evaluations in the world'' would provide ``no more useful signal,'' due to complexity and evaluation awareness (US36). 
 
Importantly, participants noted that the inability argument currently provides people's ``sense of security,'' and once it is gone the situation becomes ``much more precarious'' (UK19). It remains ``very unclear [\ldots{}]  whether we should expect safety cases to be used in the future'' (UK22).

These concerns challenge the assumption that evaluation methodology keeps pace with development, and that evaluations can continue to produce a trustworthy signal as models scale (Table~\ref{tab:assumptions}).

\subsection{Evaluation-Based Governance May Prove Counterproductive}
\label{sec:counterproductive}
 
\paragraph{Path dependency and agenda setting.}
Participants indicated that evaluations may have become the 
dominant governance tool partly because AI companies were 
already using them for capability development and marketing 
(US37, UK17; see also \citealt{Ren2024}; 
\citealt{BirhaneKalluriCard2022}). As one evaluation developer put it, 
``evaluations used to be indicators for a field to measure 
its own progress against abstract but shared goals, such as 
language understanding [\ldots{}] I don't think they are 
suited to predict real world performance or safety, and I 
actually don't think that we've made many efforts to try to 
change that'' (I13). Participants argued that developers' 
voluntary adoption of evaluation-based governance could have 
narrowed the regulation conversation to a test-and-mitigate 
paradigm (US27, US29; see also \citealt{Wei2024}; 
\citealt{Anderljung2023}), an arrangement that ``looks 
pretty good from developers' perspective because it's better 
than harder regulation and in practice constrains their 
behavior basically not at all'' (US27).
 
\paragraph{Permission structure.}
Several participants went further, arguing that the governance framework built around evaluations could absorb political energy that would otherwise push toward stronger regulatory interventions (UK17, US29, US24, EU9). A2 described voluntary commitments as ``kind of just good PR and you don't have to do much for it. And you don't have to respect it at all.'' UK17 called evaluation-based governance ``essentially permission notes to do something very dangerous'' through ``an internal voluntary mechanism [with] no legal teeth behind it.'' US29 connected this to the view that the path to safe AI runs through building increasingly capable models and iterating on safety as problems emerge (see \citealt{OpenAISafetyAlignment2025}), arguing that ``this whole paradigm [of evaluation-based governance] is partly to justify the development'' of increasingly capable models. As US29 goes on to note, ``this is science, this is responsible, how could it be bad to know more -- but it is bad to develop the dangerous thing, even if your intent is to learn more.'' The quantitative appearance of evaluations could reinforce this, as ``the image of evaluations is pretty sciencey and rigorous and systematic, even when they are very vibey and ad hoc'' (P15). Participants noted that setting thresholds could also shift the burden of proof onto those claiming risk, as ``the underlying assumptions when you set evaluations thresholds is to say, under this threshold, it's fine [\ldots] it hides the fact that we don't really know what we're doing'' (EU9). This pattern may resemble dynamics documented in other industries, where companies publish metrics that create the impression of accountability without reducing harm \citep{OreskesConway2010, MetcalfMossBoyd2019}. US24 captured the downstream consequence of this dynamic: \begin{quote}``The policymakers have been like, `oh, well, you're just evaling the models. We've got a plan. Everything will be fine. Look, you've got your [evaluation-based governance framework].' [\ldots] I really don't want to create this impression among the policymakers [\ldots] I want people to be like, oh no, there's no plan currently. We're doing this evaluations thing because it's the only thing we know how to do.''\end{quote} One former researcher at a frontier AI developer captured this starkly: ``if you do the eval and the eval is like the model can't do this thing, you're like good, I tried and I feel safer because of it [\ldots{}]  it's comforting. It's false.'' (I11)
 
\paragraph{Normalizing risks.} The continuous publication of incrementally alarming results could also erode the alarm needed to generate political will. P15 described catching themselves experiencing this: ``it's possible if I went back in time and talked to 2022 me they would be really freaking out about [today's] system card.'' Because the governance framework responds to specific capability thresholds rather than questioning the overall trajectory, each individual result could be treated as manageable even as the cumulative picture grows more alarming(see also \citealt{Williams2025}).

Nonetheless, most other participants who raised concerns 
still considered evaluations positively. US34 noted that 
``humanity is almost certainly better off in some sense 
with the evaluations than without them.''
 
The counterproductive dynamics described here do not map 
onto a single assumption failure. Instead, they suggest that 
the governance paradigm built around evaluations could 
undermine the political conditions needed for the other 
assumptions to hold (Table~\ref{tab:assumptions}).
 
\section{Strengthening Evaluation-Based Governance}
 
Alongside the faults with evaluations, participants described conditions that could be needed for evaluation-based governance to succeed. These conditions map onto the assumption chain documented in Table~\ref{tab:assumptions}.
 
\paragraph{Improving how evaluation results are communicated.}
The most widely raised condition was making evaluation results more legible to non-technical audiences (8 of 37 participants). US30 argued that ``the way evals are normally presented right 
now are not that digestible for decision-makers'' and that 
the field needs ``people doing this translational work'' to 
bridge the gap between technical outputs and policy audiences. Six stressed that evaluations should not be treated as the sole means of communicating AI risks; the most commonly named complements were dangerous capability demonstrations (4), which UK17 argued are more persuasive because ``you can very 
clearly understand the implication,'' and real-world incident data from post-deployment monitoring (3). On the technical side, participants suggested maturing evaluation best practices (3) and developing propensity evaluations alongside capability evaluations (3).
 
\paragraph{Building institutional capacity to respond to evaluation results.}
Seven participants argued for greater investment in safeguards and mitigations (the ``then'' of if-then commitments) since mitigation readiness was identified as one of the more fragile assumptions (Section~\ref{sec:noaction}). Six emphasized building technical expertise within government to improve regulator response and scrutiny of potential agenda-setting capture. UK21 drew a parallel with financial regulation, 
arguing that effective AI oversight requires ``a lot of 
technical understanding, as it seems like financial regulators 
do''. Seven argued that governance work should take priority over technical evaluation work. US23 put this directly: ``I would 
put the institutional incentives before the technical, because 
I think if you have the institutional incentives then the 
funding and the talent and the time to increase the quality 
of the technical side of things would then come.''
 
\paragraph{Establish a rigorous third-party evaluator ecosystem.}
Eleven participants argued that external parties need deeper 
access to AI models for evaluation-based governance to 
succeed. US35 
argued that evaluations could ``still be pretty impactful'' 
if ``third-party orgs are still getting access to the 
models and are not six months behind the frontier,'' but 
warned that ``this falls apart if the labs start not 
releasing their models or keeping them internal for 
extended periods of time.'' Several participants 
argued that a legislated auditing framework could address 
both access and independence, since evaluators who depend 
on voluntary partnerships with developers may find it 
difficult to present critical findings without risking 
those partnerships (see Section~\ref{sec:labcontrol}).
 
\paragraph{Shifting strategic priorities.}
Three advocated for building the institutional machinery 
needed for a slowing of AI capability improvement, arguing 
that if evaluations lose their usefulness as models grow 
more capable, the only remaining lever may be to slow capability 
development itself.

\section{Discussion and Limitations}
 
Our interviews were conducted between February and April 
2025, providing over a year of subsequent developments 
against which to assess participants' concerns. Three 
recent developments suggest these concerns may have been well-placed.

\paragraph{Evaluation inadequacy amidst capability progress.}
Several participants warned that evaluations could lose 
informativeness as models outpace the methodology available 
to assess them (see Section~\ref{sec:diminishing}). This 
concern now appears to have empirical grounding. In an April 2026 release, Anthropic stated that ``Claude Opus 4.6 has saturated most of our automated evaluations, meaning they no longer provide useful evidence for ruling out [higher risk levels of] autonomy,'' and that ``we will likely discontinue [these evaluations] going forward,'' with the safety determination resting ``primarily on an internal survey of Anthropic staff'' \citep{AnthropicOpusSystemCard}. This could reflect a central tension, where evaluations are becoming more 
load-bearing in governance frameworks at the same time they may be becoming 
less informative, and as systems could be becoming riskier.
Further, this could leave 
developers as both the judges of when their own evidence 
base has degraded and the decision-makers about what 
follows from that judgment.

\paragraph{Voluntary commitments may have weakened under competitive pressure.}
Several participants predicted that developers would roll back safety commitments when faced with racing dynamics, expecting them to say ``this other lab is racing too much [\ldots{}]  we also have to race'' (US24), or invoke geopolitical competition as justification. In February 2026, Anthropic's renewed Responsible Scaling Policy (i.e.\ their evaluation-based governance framework) dropped its prior commitment to unilaterally pause development if mitigations could not sufficiently reduce risk, replacing it with ``competitor-contingent commitments'' that condition some safeguards on whether competitors adopt matching standards \citep{AnthropicRSPv3}. Anthropic justified this by arguing that unilateral slowdowns could result in ``a world that is less safe'' because ``the developers with the weakest protections would set the pace'' \citep{AnthropicRSPv3, Karnofsky2026}. Both Anthropic and OpenAI have also publicly cited Chinese AI development as a reason to accelerate capability development \citep{Anthropic2028Leadership, Amodei2025DeepSeek, OpenAI2025}. Such arguments are not unreasonable; a unilateral pause by a safety-conscious lab 
may not reduce overall risk if less cautious competitors fill 
the gap \citep{Williams2025}.
Nevertheless, the shift illustrates how competitive pressure could 
erode commitments in ways participants anticipated. These developments are consistent with participants' concerns that if-then commitments ``can kind of just change [\ldots{}]  at any time'' (A6) and that geopolitical tensions could be used to justify continued development, potentially undermining the assumptions that evaluation results are acted upon (Table~\ref{tab:assumptions}). 
 
\paragraph{Evaluation awareness may be undermining evaluations' informativeness.}
Many participants worried that models would develop the ability to detect evaluation conditions and alter their behavior accordingly (see Section~\ref{sec:diminishing}). This concern now appears to have support. For instance, in Anthropic's recent Claude Opus 4.6 System Card, an external evaluator reported that during preliminary testing they ``observed high levels of verbalized evaluation awareness'' and concluded they could not provide a formal assessment without substantial further experiments \citep{AnthropicOpusSystemCard}. OpenAI's GPT-5.5 and Meta's Muse Spark showed similar patterns, with high rates of verbalized evaluation awareness, and Meta acknowledging that ``this raises the risk that the model could be behaving well on [\ldots{}] alignment evaluations because it believes it is being evaluated'' \citep{OpenAIGPT55, MetaMuseSpark}. If capability evaluations no longer reliably correspond to underlying capabilities, the empirical foundation for if-then commitments, red lines, and regulatory risk classification could be undermined.
 
Taken together, these developments suggest a pattern many participants anticipated, where the institutional and political conditions on which evaluation-based governance depends could be eroding faster than the technical foundations are maturing. The question may not be whether evaluations produce useful information, but whether the governance frameworks built around them can translate that information into reduced risk when the actors who produce the evidence also decide what it means and what follows from it.

\paragraph{Limitations}
 
This work has several limitations. Our sample of 37 participants was recruited through purposive and snowball sampling, which likely over-represents individuals already engaged with questions about evaluation limitations. Perspectives from senior commercial leadership at frontier AI companies, and from non-Western jurisdictions, are underrepresented. Interviews were conducted between February and April 2025, during a period of rapid change in AI governance, and some assessments may already require updating. The interview protocol's structure---moving from theories of change to assumptions to criticisms---may have primed participants toward identifying failure modes, though it also included explicit questions about which theories of change participants found most compelling. Our analytical reconstruction of assumptions (Tables~1 and 2) involved interpretive judgment; different analysts could plausibly organize the same material differently, though we followed established interpretive coding methods \citep{Saldana2021} and both coders reviewed the final mapping. Finally, this study examines evaluation-based governance as a paradigm rather than the technical quality of specific evaluations. Advances in evaluation methodology could address some measurement limitations, though most participants expressed skepticism about institutional and political challenges that technical progress alone would not resolve.

\section{Conclusion}
 
Evaluation-based governance occupies a central but precarious position in current approaches to managing risks from frontier AI. Our findings suggest this reliance could rest on foundations which experts embedded in the paradigm themselves doubt. Participants identified three pathways through which evaluations could reduce risk – making risks concrete, anchoring governance commitments, and shaping safety incentives – but each depended on assumptions about evaluation quality, honest communication of results, and appropriate institutional responses that participants widely questioned. Several of these concerns have already begun to materialize, such as voluntary commitments weakening under competitive pressure, evaluation awareness undermining results, and evaluation saturation leaving developers reliant on weaker evidence bases for risk decisions. Some experts even concluded that the evaluation-based governance paradigm was unfit for increasingly capable models, given these weaknesses. None of this means evaluations should necessarily be abandoned -- participants identified ways to strengthen the paradigm, including mandatory independent auditing with sufficient model access, binding regulation specifying which results trigger which responses, and greater investment in mitigation readiness. Whether these conditions can be met given the pace of AI development and the structural incentives facing frontier AI companies, however, is an open question.

\section*{Ethics Statement}
 
We followed established best practices for protecting interviewee confidentiality and de-identifying interview data \citep{SaundersKitzingerKitzinger2015}. All participants received a participant information sheet in advance and provided verbal informed consent prior to recording. Participation was voluntary and uncompensated, participants could decline any question or withdraw their data prior to anonymization, and they were informed that withdrawal would not be possible after anonymization.
 
Given that many participants work in a small professional community where views on AI governance can have career consequences, we took particular care with anonymization. Participants were assigned a random ID paired with a general occupation-group descriptor (e.g.\ ``an evaluations researcher at a US nonprofit''), confirmed during the interview. Identifying details such as employer, name, and specific role were redacted from transcripts, while references to third-party organizations and public figures were preserved. Recordings were stored in an encrypted folder for less than 48 hours before deletion; at no point were recordings or identifiable data input to AI systems other than a local transcription tool.

\section*{Positionality Statement}
 
The first author conducted this research as a scholar at MATS Research, supervised by a faculty member at Universit\'{e} de Montr\'{e}al. This positions the first author within the AI safety research community, the same community from which many participants were recruited. This proximity facilitated access to a specialized expert population and enabled rapport on technically complex and institutionally sensitive topics. It also means the first author shares some professional networks, assumptions, and concerns of the participants. We mitigated interpretive bias through a structured interview protocol refined through pilot interviews, independent coding by a second coder, and inclusion of participants across four distinct stakeholder groups. Nevertheless, readers should be aware that the study was designed and led by a researcher invested in the question of whether evaluation-based governance can succeed, and this investment likely shaped both the questions asked and the analytical emphasis placed on points of failure.

\section*{Adverse Impact Statement}
 
This research carries risks alongside its intended benefits. There is a risk that our emphasis on shortcomings is read as an argument that evaluations are useless, or that evaluation-based governance has already failed. We do not claim either. Most participants considered evaluations net positive, and many conditions they identified for strengthening the paradigm remain achievable. Our findings are intended to inform efforts to improve evaluation-based governance, not to foreclose them.
 
There is also a risk that by documenting how evaluations could function as a ``permission structure'' for continued capability development, this article may be cited selectively to oppose evaluation-based regulation. The alternative to imperfect evaluation-based governance is not the absence of governance. Finally, our purposive sample is modest in size, and we caution against generalizing these findings beyond the population from which participants were drawn.

\section*{Declaration on the Use of AI}
Large language models were used to assist with editing and
proofreading the manuscript and to verify codes. No
identifiable data were input to these systems (see Ethics Statement). The authors reviewed all AI-assisted output and take full responsibility for the content of this article.

\bibliography{references}

\clearpage

\appendix
\onecolumn
 
\section{Participant IDs}
\label{app:participants}
 
\begin{center}
\small
\captionof{table}{Overview of interviews, with interview IDs and descriptions of experts' roles.}
\label{tab:participants}
\begin{tabular}{llp{12cm}}
\toprule
\textbf{ID} & \textbf{Type} & \textbf{Role} \\
\midrule
A1 & Academia/research & A PhD student in technical AI governance \\
A2 & Academia/research & A PhD student in AI and society \\
A3 & Academia/research & A UK postdoc doing evals research \\
A4 & Academia/research & A PhD student in AI ethics \\
A5 & Academia/research & A researcher in robustness of evaluations \\
A6 & Academia/research & A PhD student in AI evaluations and alignment \\
A7 & Academia/research & A PhD student in AI evaluations and alignment \\
\midrule
EU8 & EU civil society & A technical AI risk management expert at an EU think tank \\
EU9 & EU civil society & A technical AI risk management expert at an EU think tank \\
EU10 & EU civil society & An AI governance researcher \\
\midrule
I11 & Industry & An evals developer, formerly a researcher at a frontier lab \\
I12 & Industry & An evals developer at a frontier lab \\
I13 & Industry & An evals researcher at a frontier lab \\
I14 & Industry & A former policy lead at a frontier lab \\
\midrule
P15 & Philanthropy & A grantmaker on technology policy at a philanthropic organization \\
\midrule
UK16 & UK civil society & A technology policy researcher at a UK think tank, formerly at UK government \\
UK17 & UK civil society & A policy expert for an AI safety advocacy group \\
UK18 & UK government & A former technical adviser for the UK government \\
UK19 & UK government & A technical adviser for the UK government \\
UK20 & UK NGO & An evals researcher at a nonprofit \\
UK21 & UK NGO & A policy researcher at a nonprofit \\
\midrule
US22 & US civil society & An AI governance researcher at a US think tank \\
US23 & US civil society & An AI technical governance researcher at a US think tank \\
US24 & US civil society & An AI technical governance researcher at a nonprofit \\
US25 & US civil society & An AI technical researcher at a US think tank \\
US26 & US civil society & An AI technical governance researcher at a US think tank \\
US27 & US civil society & An AI technical governance researcher at a US think tank \\
US28 & US civil society & An AI governance researcher with cybersecurity expertise \\
US29 & US civil society & An AI safety advocate \\
US30 & US civil society & An AI safety advocate \\
US31 & US government & An AI standards expert formerly in the US executive branch \\
US32 & US government & An evals researcher with experience in the US government \\
US33 & US NGO & An evals researcher at a nonprofit \\
US34 & US NGO & An evals researcher at a nonprofit \\
US35 & US NGO & An evals researcher at a nonprofit \\
US36 & US NGO & An evals researcher at a nonprofit \\
US37 & US NGO & A technology policy researcher at a nonprofit \\
\bottomrule
\end{tabular}
\end{center}

\clearpage
\section{Pathway-Specific Assumptions}
\label{app:pathway}
 
\begin{center}
\small
\captionof{table}{Pathway-specific assumptions underlying evaluation-based governance.}
\label{tab:pathway}
\begin{tabularx}{\textwidth}{c >{\raggedright\arraybackslash}p{4.8cm} X}
\toprule
\textbf{\#} & \textbf{Assumption} & \textbf{Key challenges identified by participants} \\
\midrule
\multicolumn{3}{l}{\textbf{Wake up: evaluations make risks concrete to decision-makers}} \\
\midrule
W1 & Target audiences possess sufficient prior context to interpret evaluation evidence as indicating risk & Audiences without existing frameworks for AI risk may dismiss evidence as speculative or cherry-picked (\S\ref{sec:labcontrol}) \\
\cmidrule(l){1-3}
W2 & The mobilizing effect of evaluation results is not eroded by normalization or repeated exposure & Incrementally alarming results could desensitize audiences; continuous publication may normalize the trajectory rather than sustain alarm; weak mitigations could prevent the warning shots needed to generate political will (\S\ref{sec:counterproductive}) \\
\midrule
\multicolumn{3}{l}{\textbf{Commitments and regulation: evaluations anchor governance frameworks}} \\
\midrule
R1 & Capability thresholds can be defined meaningfully in advance and are not set at lab discretion & Thresholds may contain subjective wording that preserves lab flexibility; no consensus exists on which capability levels warrant which responses (\S\ref{sec:labcontrol}) \\
\cmidrule(l){1-3}
R2 & Capability gains are sufficiently gradual for governance mechanisms to keep risk levels acceptable & Automated AI R\&D could produce discontinuous capability jumps that outpace the evaluation cycle (\S\ref{sec:diminishing}) \\
\midrule
\multicolumn{3}{l}{\textbf{Safety culture: evaluations shape developer incentives}} \\
\midrule
S1 & The evaluation process compels engagement with risk rather than procedural compliance & Developers may perform the minimum required to satisfy internal or external requirements without substantive investment in the process (\S\ref{sec:labcontrol}) \\
\cmidrule(l){1-3}
S2 & Evaluation documentation is both discoverable by external stakeholders and consequential for lab decision-making & No legal framework currently makes documentation consequential; reporting conventions are absent; developers control which results are disclosed and in what form (\S\ref{sec:labcontrol}) \\
\bottomrule
\end{tabularx}
\end{center}
 \twocolumn

\clearpage
\section{Interview Protocol}
\label{app:protocol}
 
\subsection*{Section 1: Background and Expertise}
 
\begin{enumerate}
\item Briefly, what is your background in AI governance, evaluations, or policy?
\item How do you define AI evaluations?
\end{enumerate}
 
\subsection*{Section 2: Theories of Change for AI Evaluations}
 
\begin{enumerate}\setcounter{enumi}{2}
\item What has made evaluations become more popular in the last few years in AI governance, in your view?
\item Some argue that the AI safety community might be overinvesting in evals at the expense of more effective strategies. What is your view?
\item What `theory of change' for evals do you find most compelling (or plausible) for reducing societal-level AI risk?
\begin{itemize}
\item \textit{Possible follow-up:} Some say `evals alert decision-makers' or `evals incentivize safety culture in labs.'
\item \textit{If `motivate governance':} Do you see any alternative approaches---such as public demonstrations, model organisms, or `warning shots'---that might achieve the same or better outcomes at lower cost? Why or why not?
\end{itemize}
\item What non-technical conditions (e.g.\ governance, policy enforcement, industry culture) do you see as crucial for this theory of change, and where do you see the biggest fragilities?
\begin{itemize}
\item How plausible do you think these assumptions are?
\item How necessary are they?
\item How could they fail?
\end{itemize}
\item Can you describe a likely scenario in which an AI evaluation successfully prevents a catastrophe? What factors would need to align for this to happen?
\item What signals would indicate to you that evals should be deprioritized in governance approaches? Put another way, are there worlds in which evals become less impactful, and how do we know which world we're in?
\item How do you balance improving technical robustness of evals versus strengthening governance or institutional incentives around evals? Which do you see as more lacking or critical right now?
\end{enumerate}
 
\subsection*{Section 3: Challenges and Limitations}
 
\begin{enumerate}\setcounter{enumi}{9}
\item Can you envision scenarios where even highly accurate or reliable evals fail to prevent harmful deployments? What factors would lead to that failure?
\item External factors---such as political will, geopolitics, progress in AI capabilities, presence of scheming, or industry incentives---could play a role in shaping the success or failure of evals' theory of change. Which external factor is most influential for whether evals are useful, in your view?
\item How do you see the role or conduct of evals changing in the next 2--3 years, if at all? (Who conducts them; how; for what purpose.)
\item Are there any negative effects of conducting evaluations that seem underappreciated or important to highlight?
\item Could evals themselves inadvertently cause harm---for example, by revealing dangerous capabilities to malicious actors or fueling an AI arms race? What safeguards might prevent this?
\end{enumerate}
 
\noindent\textit{Additional questions for technical researchers:}
 
\begin{enumerate}\setcounter{enumi}{14}
\item How would you characterize the current `science of evals'? (1) Definition; (2) what is its goal; (3) how confident are you we'll reach this goal (at all, and within 3 years)?
\item What fundamental limitations of evals are underappreciated or most important to highlight?
\item What are evals good for?
\item How much more difficult are control or propensity evals? How confident are you that these will be sufficient within 2--3 years? What unique technical challenges do they present? Do they present any unique governance challenges (e.g.\ more subjective or contentious)?
\end{enumerate}
 
\subsection*{Section 4: Governance Questions}
 
\begin{enumerate}\setcounter{enumi}{18}
\item Are there misconceptions you think people in government have about evals, or misunderstandings that you've come across?
\item Some worry that a strong focus on making evals technically robust could create a false sense of security among AI developers and policymakers. The argument is that having reliable evals signals that evals alone can assure safety. How concerned are you about this?
\end{enumerate}
 
\subsection*{Likert-Scale Survey}
 
\noindent\textit{Instructions:} For each statement, please indicate your level of agreement from $-2$ (strongly disagree) to $+2$ (strongly agree), or ``I don't know.'' Feel free to elaborate if you wish. Please note these statements are not an expression of the interviewer's stance. If the wording is vague, think of it as your interpretation of the wording (e.g.\ for `adequate,' `reliable,' or `appropriate,' it is whatever you would describe as `adequate,' `reliable,' or `appropriate'). The intent is to capture your priors.
 
\begin{enumerate}
\item It is technically possible to develop evaluations that adequately measure a model's risks.
\item Current evaluation methods can adequately measure current model risks.
\item If AI capabilities advance quickly, evaluations will become significantly less reliable.
\item Evaluations can give a fair indicator of the most serious risks from AI systems.
\item In the next two years, I expect evals will be significantly depended on in AI governance.
\item Evaluations function primarily as warning signs.
\item Evaluations function primarily as safety assurance.
\item Today, I trust that leading AI labs would take appropriate action if evaluations revealed serious risks.
\item Today, I trust leading AI labs to conduct thorough and honest evaluations of their models.
\item Leading AI labs will use evaluations primarily to demonstrate safety rather than identify risks.
\end{enumerate}

\end{document}